# Research on Jing Dong's Self-built Logistics Based on Technology Acceptance Model

Yunsheng Wang[1, *, †] and Jiaxuan Zhao[2, †]

[1]College of Mathematics and Statistics, Shenzhen University, Shenzhen, 518060, China

[2]Business school, Shandong University, Weihai, 264209, China

* Corresponding author: 2019193031@email.szu.edu.cn

†These authors contributed equally

**Abstract.** Today is a time of rapid e-commerce development, and Jing Dong China's e-commerce giant has taken its place in the highly competitive industry with its self-built logistics system. This paper analyzed the impact of Jing Dong's self-built logistics system characteristics on user satisfaction and continuous use intention by using the Technology Acceptance Model as the theoretical framework. This paper collected 295 valid samples using a questionnaire survey, all the respondents are users and potential users of Jing Dong from mainland China. The empirical results of data analysis showed that marketing information quality, logistics system quality, and logistics service have significant effects on the perceived usefulness of Jing Dong's self-built logistics, while only marketing information quality and logistics system quality have significant effects on the perceived usefulness of self-built logistics among the self-built logistics system characteristics dimensions. Additionally, the willingness to continue using a product and user satisfaction were both directly and significantly impacted by perceived usefulness, perceived ease of use had an indirect impact on users' willingness to continue use by affecting perceived usefulness and user satisfaction, and user satisfaction has the most significant impact on users' continuous use of Jing Dong shopping and using Jing Dong self-built logistics.

## 1. Introduction

### 1.1 Research Background

Online shopping is becoming the most preferred method of shopping in an era of rapid growth in the transport and traffic sectors. Due to the variety of options and convenience, the competition between domestic e-commerce platforms in China led by Taobao, T-mall, Jing Dong, and Pinduoduo has become more and more intense. In the maturing industry competition, user traffic has become the key point of competition for each company. The major e-commerce platforms have invested large amounts of human and material resources to attract and retain users. As one of the head e-commerce platforms in China, Jing Dong Mall, unlike the cooperative logistics model of most e-commerce platforms, has pioneered a self-built logistics system to meet the higher timeliness and packaging integrity requirements of electronic products and large household appliance distribution. Jing Dong does so because third-party logistics companies have difficulty guaranteeing the quality of transportation and have high delivery fees, which seriously affects consumers' sense of service experience and shopping enthusiasm. The efficient integrated supply chain model of self-built logistics has enabled Jing Dong Mall to grow rapidly, while Jing Dong has made self-built logistics the focus of marketing and promotion for supply chain optimization and to maximize its advantages, attracting and retaining a larger number of users for Jing Dong.

### 1.2 Research Gap

Most of the studies on Jing Dong's self-built logistics are about Jing Dong's self-built logistics itself, such as the history of its establishment, how to expand it and its operation mechanism, for example, Li Danli and Sun Jinhua have sorted out Jing Dong's self-built integration operation mechanism [1]. Most of the studies on the advantages of Jing Dong's self-built logistics also stay at the qualitative level, for example, based on the five forces paradigm, Ding Li carried out a qualitative investigation on Jing Dong's self-built logistics competitiveness [2]. However, the impact of Jing Dong's self-built logistics on users' behavior has been little studied with data support. The original intention of Jing Dong's self-built logistics is to provide convenience for users, optimize user experience, attract users and retain them for Jing Dong, but few studies have been conducted to support the impact of Jing Dong's self-built logistics on user behavior, few models have also been applied to analyze how Jing Dong Mall's self-developed logistics system affects user satisfaction as well as continued usage.

### 1.3 Fill The Gap

Therefore, how does Jing Dong Mall build its logistics and implement supply chain integration as a key marketing strategy? To what extent does Jing Dong build its logistics to attract and retain users? In order to explore these two questions, this paper is based on the Technology Acceptance Model (TAM) and uses questionnaires to investigate the subjective perceptions of users and their willingness to use Jing Dong's self-built logistics system. This paper explores the mechanism of Jing Dong's self-built logistics on users' satisfaction and willingness to continue using Jing Dong, and analyzes the effect of perceived usefulness and perceived ease of use. This paper establishes a link between the characteristics of Jing Dong's self-built logistics and Jing Dong users' willingness to continue shopping at Jing Dong through a technology acceptance model, which is innovative and broadens the research perspective, and provides suggestions and countermeasures to further optimize and improve Jing Dong's self-built logistics system, realize an integrated supply chain, to attract and retain users.

## 2. Literature Reivew

Pursuing quality and price, modern e-commerce focuses on improving user experience and ensuring users' continuous willingness to use. In recent years, Jing Dong's self-built logistics system has relied on its regional supply chain integration to make up for the shortcomings of third-party logistics and has been recognized by more users. This paper focuses on the issue of Jing Dong's self-built logistics for users to grow their willingness to use continuously, firstly, we understand the current research progress and status of Jing Dong's self-built logistics, forming an integrated supply chain, and sort out the factors influencing users' willingness to use products continuously. This paper provides reference and references for this paper.

### 2.1 Research on Jing Dong's Self-built Logistics

As a leading domestic e-commerce enterprise, Jing Dong Mall sells products of many famous brands online to a wide group of users nationwide. From the perspective of Jing Dong itself, According to Liu Qiangdong, in the e-commerce competition of today, user experience is rapidly coming into focus. One of the major components of user experience is delivery speed, and only independent delivery can better understand user experience. Only tailored logistics can enhance user happiness and the customer experience [3]. Therefore, Jing Dong, an e-commerce company, has developed its logistics strategy, optimized its supply chain, interfaced directly with consumers rather than reaching them through third-party logistics, and obtained high feedback on user satisfaction ratings with its unique system of self-built logistics.

Self-built logistics, using its regional supply chain integration, has four major advantages over other e-commerce platforms that choose third-party logistics companies: first, it ensures the timeliness and safety of logistics services, second, it ensures business stability during special periods, third, it increases consumer recognition, and fourth, the warehouse vehicles have an advertising effect [4]. Each of the four advantages significantly improves the efficiency and quality of logistics services to different degrees, which enables users to get more satisfactory and high-quality services, and then realize for users to continue to use Jing Dong Mall for shopping.

## 2.2 Research on The Factors of Users' Willingness to Continuous Use

To investigate the elements that influence users' willingness to use continuously needs to be considered from two aspects, on the one hand, user attraction, and on the other hand, user retention. This paper takes the literature on user experience research as the basis of factors influencing user attraction, and the literature on user stickiness as the basis of factors influencing user retention.

### 2.2.1 User Experience Research

User experience is generally defined in academic research as the response and cognitive image of people to a product, service, or system that they use or expect to use, and includes not only the user's interaction with the product while using it, but also all the feelings (including physical and psychological reactions) before and after the use of the product. Scholars Hassenzahl and Tractinsky, on the other hand, define user experience as all the results produced by the user's internal states (tendencies, expectations, needs, motivations, emotions, etc.) and a design system with certain characteristics (purpose, usability, functionality, complexity, etc.) in a specific interaction context [5].

Hassenzahl has proposed a model called the user experience model, which explains the process of forming user experience and specifies the important factors affecting user experience: when users use a product, they initially perceive the product features (content, functionality, interaction, presentation, etc.), and then they construct the product character in their minds according to their expectations. Hassenzahl divides the product character into two categories, namely practical attributes, and enjoyable attributes, and finally, based on their perception of the product characteristics, users make their judgments about the product, i.e., whether the product is attractive, whether using the product has a positive emotional outcome for them, and whether they increase the frequency and duration of their utilization of the product or the system. The verdict of the user's judgment of the product constitutes the content of the user experience [6].

Combined with the research content of this paper, whether users can improve the user experience when using Jing Dong shopping with Jing Dong's self-built logistics system depends on the characteristics of Jing Dong's self-built logistics system and whether it can meet users' expectations of a convenient pickup, timely arrival, perfect after-sales, and other package completeness.

### 2.2.2 User Stickiness Research

From the point of view of consumers, LIN L. et al. assert that user stickiness is a term that describes how long visitors stay on a website after attending a meeting or overcoming a special period [7]. GalZauberman equates website locking with stickiness and approximates stickiness to consumer loyalty on a website, which can be specified as consumers' future repeat visits [8-9]. Johnson, E.J. argues that general scholars suggest that stickiness is important and a significant factor in favor of e-commerce retailers [10]. Khoshoie originally defined stickiness as the highest frequency, depth, and duration of use by users to a live website [11].

In customer relationship management, many scholars believe that customer persistence depends on customer satisfaction and customer satisfaction depends on the level of expectation confirmation [12]. Limayem proposed that user habits are a negative moderating variable in the connection involving persistent intention to use and persistent behavior and that the moderating effect.

The moderating effect will gradually increase over time [13]. Based on the overall e-commerce product, Ying Chen mainly combines the theory of loyalty with the background of B2C websites and the characteristics of their shopping experience to build a corresponding research model to study the user experience brought by the overall e-commerce product [14]. In constructing a study model of customer loyalty for software operation services, Miao Shengtao used factors such as trust, reliability, and ease of use to explore their relationship with customer satisfaction and customer loyalty [15]. Jun Xue and Qing Zhao et al. took the factors influencing consumer stickiness behavior in online shopping in front of the net as the entry point, and explored their relationship with consumer stickiness based on the IS-ECM model (Information System Continuity of Use Model) proposed by Bhattacherjee in information systems science, using factors such as usefulness, satisfaction, and heart flow as antecedents, and concluded that consumer perception factors have an impact on stickiness through mediating variables conclusions such as the influence of consumer perceptions on stickiness through mediating variables [16].

According to the existing studies, the focus and research methods for user stickiness vary, but most of them include trust and satisfaction in the study for analysis and measurement, which provides a reference for this study. Combined with the content of this study, the trust degree and satisfaction of Jing Dong users for Jing Dong logistics can be used as exploration factors.

**2.3 Model Basis**

Davis first proposed the Technology Acceptance Model (TAM) in 1989, based on the theory of rational behavior, which he initially used to study the principles of user acceptance of information systems to explain the phenomenon of widespread acceptance of computers in the prevailing environment and to find mechanisms and causes for this phenomenon to work. The two primary variables suggested by Davis form the basis of the technology acceptance model: perceived usefulness and perceived ease of use. The model framework ultimately reflects an individual's actual use of a system through these two key factors, where perceived usefulness is the degree to which people's usage of a particular system enhances their ability to perform at work, and perceived ease of use is the degree to which people find a particular system simple to use. The model framework is as follows (Figure 1).

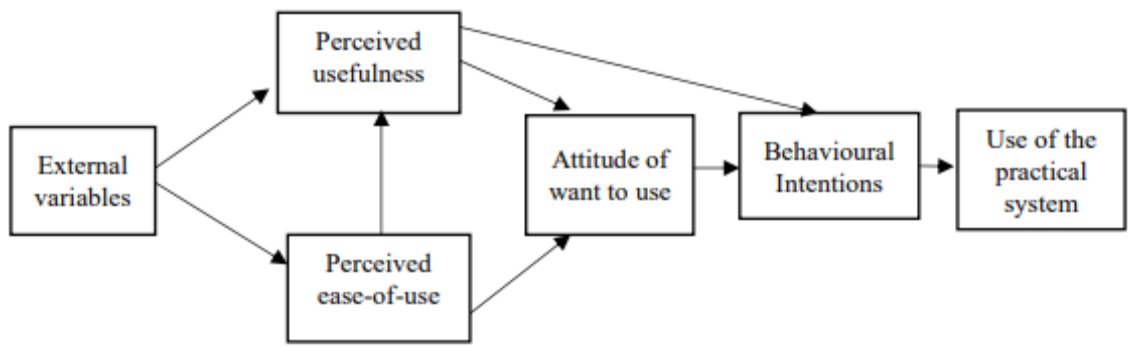

**Figure 1.** Framework

Although originally proposed to explore people's willingness to accept and use new technologies and systems, TAM is increasingly utilized in other domains, such as psychology [17]. In terms of the research of this model in the field of e-commerce, Yang Xiaomei found that the field helps to guide consumers to choose the appropriate transaction objects and tap potential consumers; Zhu Xiaodong et al. explored the key factors affecting the willingness to use cloud service platforms of e-commerce enterprises based on the integrated TAM; Zhang Mengxia confirmed through TAM that initial trust can have an impact on consumers' final decision to purchase products, etc. [18-20]. Some scholars have drawn on the use of TAM in the sphere of e-commerce to consider social e-commerce as an information system and analyze users' behavioral willingness from multiple perspectives, but few have analyzed the impact of self-built logistics on e-commerce users' continued willingness to make. Given this, this paper will use the TAM-based model as the framework for exploring this paper, in which perceived usefulness measures how much users believe using Jing Dong self-built logistics would benefit their life, while perceived ease of use measures how simple they believe using Jing Dong self-built logistics will be, to reveal the role of Jing Dong self-built logistics on users' willingness to use Jing Dong Mall consistently.

## 2.4 Literature Summary and Review

At present, the academic community is rich in research on Jing Dong's self-built logistics and the factors of users' willingness to continue to use it. From the aforementioned literature review, this paper derives the following findings: the current academic research on the impact of Jing Dong's self-built logistics on users' willingness to continue to use has been relatively abundant, which provides a rich experience and theoretical reference for the subsequent expansion of research. We found that there are mostly empirical studies in the literature, and there is no lack of studies on the direction of user satisfaction, but there are fewer studies that use the TAM model to establish the connection between the two.

In this regard, this paper mainly makes the following contributions: introducing TAM to establish the relationship between self-built logistics systems and user growth, analyzing the prospect that self-built logistics is relatively better than traditional logistics forms, which is more effective in improving user satisfaction and service evaluation, and based on the TAM concept, fully demonstrating the relationship between the two thru perceived usefulness and perceived ease of use, broadening the perspective for e-commerce It provides valuable suggestions and countermeasures to attract and retain users.

## 3.Methodology

### 3.1 Research Design

In this paper, three dimensions of Jing Dong's self-built logistics system are extracted as the main external variables that mainly affect TAM, to explain the impact of Jing Dong's self-built logistics on users' willingness to continue to use it, and the following figure depicts the model framework created in this paper (Figure 2). The research logic is that users' willingness to keep using Jing Dong Mall depends on their perceived usefulness and perceived ease of use, and external variables influence users' behavior by affecting perceived usefulness and perceived ease of use. In the following, we elaborate on each variable of the model and make the research hypothesis of this paper.

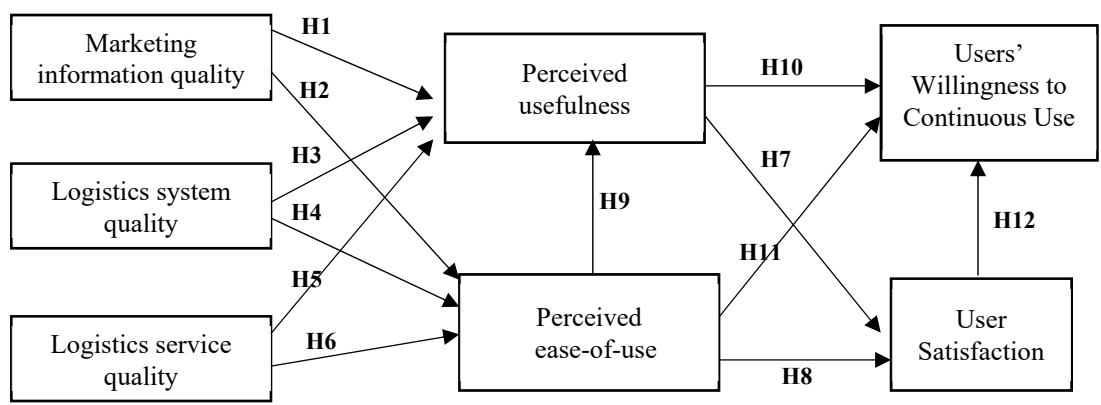

**Figure 2. Model Framework**

### 3.1.1 Jing Dong Self-build Logistics and TAM

Marketing information quality refers to the information of Jing Dong's self-built logistics provided to users by Jing Dong through operating social platforms or placing advertisements, and key features such as timeliness, accuracy, adequacy, and practicality are information quality. For Jing Dong's self-built logistics, the quality of marketing information is very important, because most e-commerce platforms cooperating with third-party logistics platforms, which will lead to the cumbersome logistics process, while Jing Dong has its integrated supply chain, self-built logistics standardized management, simplify the logistics process, and more powerful protection for large home appliances and digital products, improve timeliness and reliability; improve user satisfaction, if the marketing If the quality of information is good, then users will also be more willing to use Jing Dong self-built logistics system and choose Jing Dong Mall for shopping.

The quality of the logistics system is a measure of the quality of Jing Dong's self-built integrated logistics system itself. The quality of Jing Dong's logistics system reflects the system characteristics of logistics, such as logistics timeliness, reliability, smoothness of docking with customer service, and timeliness of logistics information update. The quality of the logistics system is the basis for the quality of marketing information and logistics service to reflect its value, so the quality of Jing Dong's self-built logistics system is the cornerstone, if the quality of the logistics system is poor, such as not timely delivery, not more detailed logistics information, logistics pages are difficult to find, etc., the user will form a poor first impression of the logistics system, and will directly lead to give up shopping in Jing Dong Mall, let alone continue to use it.

Logistics service quality refers to the effect of users using Jing Dong's self-built logistics services, such as the attitude of Jing Dong's logistics customer service, problem-solving ability, etc. Previous studies by many scholars have shown that service quality is a decisive factor affecting user satisfaction, and for Jing Dong's self-built logistics system, service quality is a must, and logistics service quality reflects the characteristics of website services, including reliability, responsiveness, and so on. Reliability shows that Jing Dong logistics customer service can provide users with on-time service, reasonable packaging, complete goods, etc.; responsiveness requires logistics service personnel to provide timely responses to users' questions, whether returns and exchanges are smooth, etc., to build users' confidence in them.

The above three dimensions of marketing information quality, logistics system quality, and logistics service quality will all affect its ease of use and usefulness. If the above three dimensions are poor, it will be difficult for users to feel the ease of use and usefulness of integrated logistics, such as the lack of timely logistics delivery and slow response time of logistics services, and its ease of use and usefulness will be greatly reduced. Therefore, this paper makes the following assumptions (the framework is shown in Figure 3).

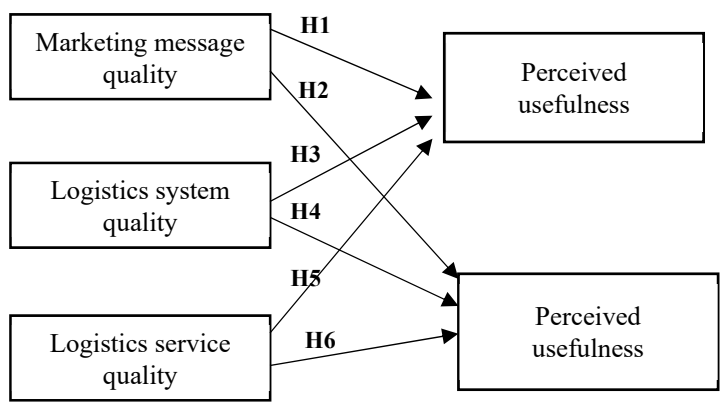

**Figure 3. Framework 2**

H1: Users' perception of the quality of Jing Dong logistics marketing information will significantly affect their perception of the usefulness of Jing Dong self-built logistics.

H2: Users' perceptions of the quality of Jing Dong logistics marketing information will significantly affect their perceptions of the ease of use of Jing Dong self-built logistics.

H3: Users' perceptions of the quality of Jing Dong's self-built logistics system will significantly affect their perceptions of the usefulness of Jing Dong's self-built logistics.

H4: Users' perceptions of the quality of Jing Dong's self-built logistics system will significantly affect their perceptions of the ease of use of Jing Dong's self-built logistics.

H5: Users' perceptions of the quality of Jing Dong's self-built logistics services will significantly affect their perceptions of the usefulness of Jing Dong's self-built logistics.

H6: Users' perceptions of the quality of Jing Dong's self-built logistics services will significantly affect their perceptions of the ease of use of Jing Dong's self-built logistics.

### 3.1.2 TAM and User Satisfaction

Many scholars assert that perceived usefulness and perceived ease of use demonstrate a strong association with user satisfaction with information systems [21-23]. If users feel that an information system is valuable, they will also be satisfied with the system and thus accept to use the information system. In a study of user satisfaction with mobile government systems based on TAM, some scholars also found that user satisfaction was primarily determined by perceived usefulness, with perceived ease of use having a secondary impact [24]. In this paper, the information system in the past research is extended to the whole Jing Dong self-built logistics system including logistics information system, therefore, the hypothesis is made in this paper (the framework is shown in Figure 4).

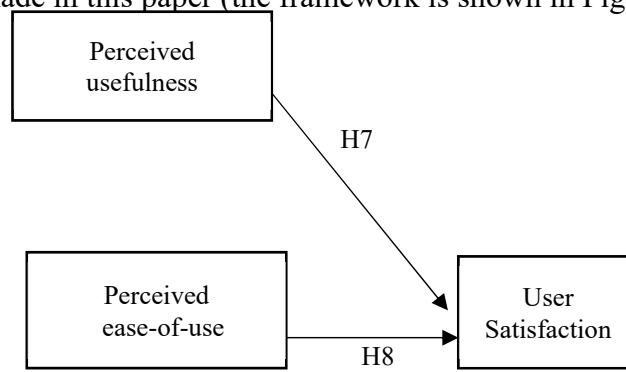

**Figure 4. Framework 3**

H7: users' perceived usefulness of Jing Dong's self-built logistics system will significantly affect users' satisfaction level.

H8: users' perception of the ease of use of Jing Dong's self-built logistics system will significantly affect users' satisfaction.

### 3.1.3 TAM and Users' Willingness to Continuous Use

Although TAM is used to explain users' initial intention to use an information technology or information system, it can actually be used to account for users' intention to use the information system afterwards. Taylor and Todd demonstrated that TAM can be applied to experienced or inexperienced users' use of the system [25]. Bhattacherjee used expectancy confirmation theory to construct the information system sustained use model, also emphasized that TAM can be used not only for initial acceptance but also to account for post-acceptance usage intentions. Moreover, the empirical findings show that perceived usefulness has a significant positive effect on user satisfaction as well as intention to continue using, respectively [26]. Therefore, this paper makes the hypothesis (framework in Figure 5) that:

H9: users' perception of ease of use of Jing Dong self-built logistics will significantly affect their perception of usefulness.

H10: users' perception of usefulness of Jing Dong self-built logistics will significantly affect their intention to continue using it.

H11: Users' perception of the ease of use of Jing Dong's self-built logistics will significantly affect their willingness to continue shopping with Jing Dong Mall

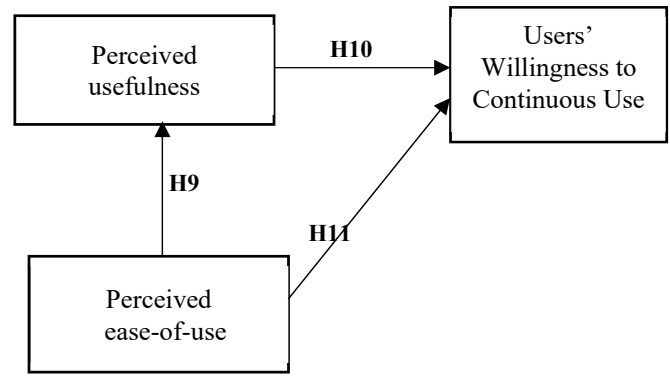

**Figure 5. Framework 4**

### 3.1.4 User Satisfaction and Users' Willingness to Continuous Use

The purpose of Jing Dong's self-built logistics system is to provide users with a higher quality shopping experience and attract more users to shop at Jing Dong Mall; therefore, the success of Jing Dong's self-built logistics depends on users' satisfaction with Jing Dong's self-built logistics and their intention to continue using Jing Dong Mall. Satisfaction with Jing Dong's self-built logistics refers to the public's psychological or emotional state, which originates from the cognitive evaluation after using it [26].

Users' intention to shop at Jing Dong Mall and to continuously use self-built logistics can be understood as repeated use of the Jing Dong e-commerce platform for shopping. The continued use depends on two aspects, first is the determination before using the website, which is a repeated behavioral mechanism, and the second is the subjective evaluation after using the website, which is a feedback mechanism [24].

This argument states that the feedback mechanism can explain the connection between user satisfaction and willingness to continue using and that user satisfaction will have a major impact on users' desire to continue using. users' willingness to use will be significantly impacted by their level of satisfaction. Applied to the study in this paper, if users feel more satisfied when they shop in Jing Dong Mall and use Jing Dong's self-built logistics, then they will shop in Jing Dong Mall next time. Therefore, the following hypothesis can be made: (The framework is shown in Figure 6)

H12: Users' Jing Dong logistics satisfaction will significantly affect their willingness to use Jing Dong Mall for shopping.

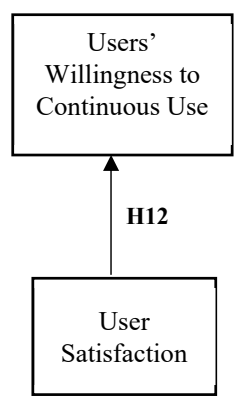

**Figure 6. Framework 5**

### 3.2 Data Collection

#### 3.2.1 Variable Measurement

To make certain that the scale's content validity, the variables in this study were measured from previous classical literature and modified in the context of Jing Dong's self-built logistics user satisfaction survey. According to the research integration model, this paper set up questions with different measures for the seven investigated variables: marketing information quality, logistics system quality, logistics service quality, perceived usefulness, perceived ease of use, user satisfaction, and user willingness to continue using, and the first draft of the questionnaire was formed after a series of language adjustments. The questionnaire was measured using a 5-point Likert scale, that is, respondents can rate themselves on a scale of 1 to 5 depending on their level of agreement with the question, with a score of 5 when they fully agree and 1 when they fully disagree. The respondents were asked to assess within 1-5 points according to their situation. To confirm that the questionnaire's content is logical and the meaning is expressed accurately, this paper conducts pre-research based on the first draft of the questionnaire. The pre-study was published online, and then the questionnaire was revised in response to the pre-study to form the official questionnaire of this paper. Table 2 displays the findings of the variables' descriptive statistics.

#### 3.2.2 Data Source

In terms of the sampling method, this paper adopts the approach of a convenience sample. The questionnaires were distributed online on the web platform Questionnaire Star from July 31, 2022, to August 9, 2022. The returned questionnaires were screened and a total of 320 questionnaires were received, of which 295 were valid, and the questionnaire efficiency rate was 92.18%, all the respondents are Jing Dong users and potential users from mainland China. Among the valid questionnaires, 127 were from men, accounting for 43.05%, and 168 were from women, accounting for 56.94%. From the sample, we can see that the proportion of males to females is equal, and the proportion of females is slightly higher. Regarding age, more than half of the sample was composed of young and middle-aged people aged 19-35, accounting for 58.64% of the total sample. In terms of education, the highest proportion is undergraduate and above, accounting for 65.08%, followed by high school/junior college/technical school education, accounting for 19.66%. In terms of monthly income, students accounted for the main questionnaire-issuing group, with a slightly higher proportion of the monthly income of RMB 3,000 and below, accounting for 29.15%. Students made up the largest share of the sample in terms of occupation (23.39%), followed by professionals, with 21.69%. (Specific information is shown in Table 1)

Table 1. Basic information about the respondent

|  | Number | Proportion |
|---|---|---|
| Gender |  |  |
| Male | 127 | 43.05% |
| Female | 168 | 56.95% |
| Age |  |  |
| Under 18 years old | 12 | 4.07% |
| 19-25 years old | 96 | 32.54% |
| 26-35 years old | 77 | 26.10% |
| 36-45 years old | 59 | 20.00% |
| 46 years and above | 51 | 17.29% |
|  | Number | Proportion |
| Education level |  |  |
| Junior High School and below | 22 | 7.46% |

| | | | |
|---|---|---|---|
| High school/junior college/technical school | 58 | 19.66% | |
| Bachelor's degree | 192 | 65.08% | |
| Master's degree and above | 23 | 7.80% | |
| Occupation | | | |
| Students | 69 | 23.39% | |
| Professionals | 64 | 21.69% | |
| Service workers | 21 | 7.12% | |
| Freelancers | 15 | 5.08% | |
| Workers | 10 | 3.39% | |
| Company employees | 58 | 19.66% | |
| Career / civil servant / government worker | 13 | 4.41% | |
| Housewife | 12 | 4.07% | |
| Other | 33 | 11.19% | |

## 3.3 Data Analysis

### 3.3.1 Confidence and Validity Analysis

Given that SPSS is a widely used statistical tool for studies using questionnaires to collect data, and that many studies using TAM also use SPSS as a data analysis tool, this paper uses SPSS software to analyze the collected questionnaire data. The test of the measurement model involves convergent validity, discriminant validity, and internal consistency of the variables. In order to examine the questioned validity of the questionnaire, we need to analyze its reliability and validity of the questionnaire. Reliability is an estimate of the degree of consistency of the measurement of the scale. We use Cronbach's α and Composite Reliability (CR) to measure the reliability of the questionnaire, and if the above two values of each variable are greater than 0.8, therefore, have good reliability, as shown in Table 2.

Convergent validity refers to the fact that questions or tests measuring the same potential trait fall on the same factor construct. In this paper, to check the questionnaire's convergent validity, we employ average variance extraction (AVE) and factor loading coefficients. In general, the convergent validity of a questionnaire is considered appropriate when the factor loading coefficient and AVE are higher than 0.5. According to the sample analysis in this paper, the above requirements are met and the specific load factor values and AVE values are shown in Table 2.

**Table 2.** Variable descriptive statistics, reliability and convergent validity tests

| variables | Title | Average value | Standard deviations | Standard Factor Load Coefficient | Cronbach α | CR | AVE |
|---|---|---|---|---|---|---|---|
| Marketing information quality | MIQ1 | 4.23 | - | 0.820 | 0.889 | 0.892 | 0.675 |
| | MIQ2 | 3.96 | 0.066 | 0.847 | | | |
| | MIQ3 | 4.19 | 0.069 | 0.721 | | | |
| | MIQ4 | 4.06 | 0.059 | 0.889 | | | |
| Logistics system quality | LSQ1 | 4.40 | - | 0.908 | 0.927 | 0.927 | 0.810 |
| | LSQ2 | 4.36 | 0.041 | 0.902 | | | |
| | LSQ3 | 4.39 | 0.042 | 0.889 | | | |
| Logistics service quality | LEQ1 | 4.27 | - | 0.805 | 0.924 | 0.925 | 0.755 |
| | LEQ2 | 4.18 | 0.057 | 0.911 | | | |
| | LEQ3 | 4.27 | 0.06 | 0.901 | | | |
| | LEQ4 | 4.28 | 0.065 | 0.854 | | | |
| Perceived usefulness | PU1 | 4.34 | - | 0.908 | 0.910 | 0.910 | 0.835 |
| | PU2 | 4.38 | 0.038 | 0.919 | | | |
| Perceived ease-of-use | PEOU1 | 4.39 | - | 0.901 | 0.926 | 0.927 | 0.808 |
| | PEOU2 | 4.30 | 0.041 | 0.905 | | | |
| | PEOU3 | 4.26 | 0.044 | 0.891 | | | |
| User Satisfaction | SAT1 | 4.27 | - | 0.898 | 0.931 | 0.930 | 0.816 |
| | SAT2 | 4.33 | 0.043 | 0.890 | | | |
| | SAT3 | 4.33 | 0.04 | 0.922 | | | |
| Users' Willingness to Continuous Use | ITCU1 | 4.20 | - | 0.894 | 0.927 | 0.928 | 0.811 |
| | ITCU2 | 4.08 | 0.043 | 0.902 | | | |
| | ITCU3 | 4.18 | 0.047 | 0.907 | | | |

Discriminant validity refers to the existence of low correlation or significant differences between the potential traits represented by the constructs and the potential traits represented by other constructs. In this study, each variable's correlation coefficient is used to assess its discriminant validity; if it is less than the square root of the related variable's AVE value, the variable is considered to have a low correlation coefficient, and it is considered to have good discriminant validity. According to the analysis of the sample in this paper, the above requirements are met and Table 3 displays the outcomes.

**Table 3.** Correlation coefficients and differential validity tests for each variable

| | MIQ | LSQ | LSQ | PU | PEOU | SAT | IOCU |
|---|---|---|---|---|---|---|---|
| MIQ | 0.822 | | | | | | |
| LSQ | 0.66 | 0.9 | | | | | |
| LSQ | 0.669 | 0.792 | 0.869 | | | | |
| PU | 0.671 | 0.864 | 0.78 | 0.914 | | | |
| PEOU | 0.681 | 0.853 | 0.829 | 0.867 | 0.899 | | |
| SAT | 0.704 | 0.838 | 0.814 | 0.89 | 0.884 | 0.904 | |
| IOCU | 0.654 | 0.704 | 0.713 | 0.752 | 0.74 | 0.784 | 0.901 |

Note: Diagonal figures are AVE square root values

### 3.3.2 Structural Model

In this paper, the structural validity of each study scale was examined using the validation factor analysis., according to Table 4, except for GIF, which was slightly below the judgment standard, all other indicators met the judgment standard, and the model fit of this study was judged to be good.

**Table 4.** Key goodness-of-fit index for structural equations

| Indicator | X2/df | GIF | RMSEA | RMR | CFI | NFI | NNFI |
|---|---|---|---|---|---|---|---|
| Judgement criteria | <3 | >0.9 | <0.10 | <0.05 | >0.9 | >0.9 | >0.9 |
| Value | 2.524 | 0.875 | 0.072 | 0.020 | 0.961 | 0.938 | 0.952 |

The results of the path analysis show that two hypotheses, H5 and H11, were not tested. In other words, the perceived usefulness of users is unaffected significantly by the quality of the logistical service; users' willingness to continue using a product is not directly and significantly impacted by perceived ease of use. The other hypotheses were all supported by the data. Figure 7 illustrates the results of the hypothesis testing.

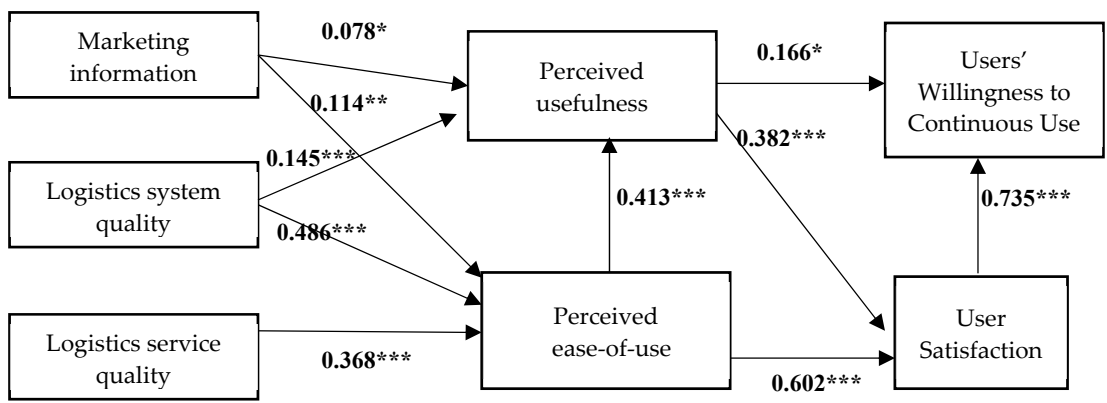

**Figure 7.** Results for hypothesis

t-statistics in parentheses  
*** p<0.01, ** p<0.05, * p<0.1

## 4. Results

This paper applies TAM as the theoretical framework and introduces three Jing Dong self-built logistics characteristics, namely, marketing information quality, logistics system quality and logistics service quality, based on previous theoretical studies, to construct a model that can explain users' continuous use behavior after Jing Dong self-built logistics. According to the study, the perceived ease of use was significantly influenced by all three of the newly introduced Jing Dong self-built logistics feature dimensions, only two feature dimensions—the quality of the logistics system and the quality of the marketing information—proved to have a discernible impact on users' perceptions of the usefulness, and logistics service quality did not prove to significantly affect perceived usefulness.

The study also discovered that perceived usefulness directly influences users' satisfaction and willingness to use a product or service in the future, however, the impact of perceived ease of use on user inclination to continue using is indirect; it does so by impacting user satisfaction, and user willingness to continue to shop at Jing Dong Mall and use Jing Dong's own logistics is most strongly influenced by user satisfaction.

## 5. Discussion

In the above summary, we can find that the three dimensions of self-built logistics system features have different impacts on users' willingness to continue shopping at Jing Dong Mall and use Jing Dong self-built logistics, and the reasons for this are closely related to the specific characteristics of the three dimensions of self-built logistics system features because it makes sense that each dimension of self-built logistics system features directly and significantly affects how easy it is to utilize the system. Only the quality of marketing information and logistics system features significantly affect the perceived usefulness because the access to information and the system itself are the main reasons that affect the user's shopping efficiency and improve the shopping experience, and the user's reliability and trust in the self-built logistics system itself mainly depend on the effectiveness of the logistics system as a whole, while perceived usefulness is not significantly influenced by logistics service quality. First, from the objective point of view, the quality of Jing Dong's self-built logistics service is not significantly different from that of third-party logistics, which leads to no obvious perception of usefulness by users; second, from the user's point of view, users have high expectations for logistics service quality, and they assert that logistics service quality should be maintained at a better level, which leads to no obvious perception of usefulness.

Previous studies by many scholars have confirmed that the core elements influencing whether users use information systems are perceived usefulness and perceived ease of use, with perceived usefulness being the decisive factor and perceived ease of use having an indirect impact mostly via perceived usefulness. This study extends the theoretical model to the whole Jing Dong self-built logistics system including information systems, and the findings reveal that customers' continuing usage of Jing Dong shopping is significantly influenced by the usability and simplicity of the self-built logistics in Jing Dong, and the usefulness of the self-built system is still a key factor in determining users' continued use of Jing Dong shopping, while perceived usefulness and user satisfaction, respectively, have an indirect impact on consumers' continuous use when it comes to perceived ease of use. Due to its effect on perceived usefulness and user satisfaction, perceived ease of use indirectly influences consumers' willingness to keep using the product. Moreover, to varying degrees, consumers' satisfaction with using the system has been linked to perceived usefulness and perceived ease of use, which has been widely confirmed in TAM, and the same has been verified in the application of Jing Dong's self-built logistics. Based on the original TMA model in the hypothesis, this research suggests a direct relationship between perceived usefulness and users' propensity to continue using, however, this relationship has not been independently validated, suggesting that in the study of Jing Dong's self-built logistics system, perceived ease of use also plays an indirect rather than a direct role in users' willingness to continue to use, which further illustrates the accuracy and wide applicability of TAM.

## 6. Conclusion

This paper explores the mechanism of Jing Dong's self-built logistics on users' willingness to continue to use Jing Dong based on TAM, and empirically analyzes the degree of influence of each variable on the willingness to continue to use Jing Dong. The research in this paper provides important implications for investigating the specific mechanism of the effect of Jing Dong's self-built logistics on users' willingness to keep using Jing Dong Mall for shopping. According to the model constructed in this paper, users' continuous use behavior is considered to be a key factor in determining the success of Jing Dong's self-built logistics and creating an integrated supply chain. If users' perceived beliefs and satisfaction with the self-built logistics system are high, then the stronger will be the users' intention to continue Jing Dong Mall shopping. Therefore, the management should pay attention to the psychological and behavioral occurrence processes of users. To promote users' continuous use of Jing Dong, managers need to strengthen the quality of marketing information, logistics system quality, and logistics service quality of Jing Dong's self-built logistics, because the three-dimensional characteristics of the self-built logistics system will affect users' behavioral intention to use and satisfaction evaluation.

In this model, user satisfaction is the primary influence on users' decision to stick with Jing Dong, which simply means that users will have a subjective evaluation of Jing Dong's self-built logistics system in general after using it, and the evaluation of whether they are satisfied with the system is the most crucial aspect that influences whether users intend to keep using it. This conclusion is particularly important because perceived usefulness is a key element in establishing users' use at the initial use stage, while in the process of continuous use, users' subjective evaluation of satisfaction plays a dominant role, from which it can be concluded that: for user groups who have not yet used Jing Dong shopping and used Jing Dong self-built logistics, they should pay more attention to the usefulness of the self-built logistics system, while user groups who have already used it, to promote their The policymakers need to pay attention to user satisfaction in order to promote their use again.

There are also shortcomings in this study. First, the number of survey samples in this study is limited, and it is dominated by highly educated groups, and the sample scope can be expanded in the future to ensure the robustness of the research findings; second, variables such as population characteristics can be further introduced into the model in the future to refine and deepen the research hypotheses and continue to improve and revise the model.